\documentclass[aps,prb,twocolumn,longbibliography,superscriptaddress,floatfix,10pt]{revtex4-2}

\usepackage[utf8]{inputenc}
\usepackage{amssymb,amsfonts,amsmath,dsfont}
\usepackage{graphicx}
\usepackage{textcomp}
\usepackage{gensymb}
\usepackage{pifont}
\usepackage{bm}
\usepackage{color}
\usepackage{accents}
\usepackage{mathtools}
\usepackage{nicefrac}
\usepackage{url}
\usepackage{multirow}
\usepackage{hyperref}
\usepackage[capitalize]{cleveref}
\usepackage{tabularx}
\usepackage{upgreek}
\usepackage{chemformula}
\hypersetup{colorlinks=true,citecolor=blue,urlcolor=blue,linkcolor=blue}

\usepackage{mathrsfs}

\newcommand{\bs}[1]{\boldsymbol{#1}}

\makeatletter

\begin{document}
\title{Magnon-dislon hybridization in magnetic insulators}
\author{Carlos Saji}
\affiliation{Departamento de Ciencias F\'isicas, Facultad de Ciencias F\'isicas y matem\'aticas, Universidad de Chile, Santiago, Chile}
\author{Nicolas Vidal-Silva}
\affiliation{Departamento de Ciencias F\'isicas, Universidad de La Frontera, Casilla 54-D, Temuco, Chile}
\author{Roberto E. Troncoso}
\affiliation{Instituto de Alta Investigaci\'on, Universidad de Tarapac\'a, Casilla 7D, Arica, Chile}
\email{r.troncoso.c@gmail.com}

\begin{abstract}
Spin dynamics in ordered magnets with topological lattice defects is investigated. Using fracton--elasticity duality, we develop an effective field theory of magnons coupled to quantized lattice dislocations (dislons) in magnetic insulators. Within this framework, an elastic gauge field mediates a nonlocal interaction between dislocations and magnetization gradients. The resulting magnetoelastic coupling gives rise to coherent magnon–dislon hybridization whose properties are dictated by dislocation topology. Screw dislocations exhibit helicity-selective hybridization and symmetry-protected dark dislon sectors, while edge dislocations generate anisotropic hybrid excitations with finite spin-precession ellipticity through the glide constraint. Our results establish dislocations as dynamical topological defects with directly observable polarization fingerprints in magnon spectra, and reveal magnon–dislon hybridization as a new route to control spin dynamics.
\end{abstract}
\maketitle

\textit{Introduction.--} Collective excitations in interacting many-body systems lie at the heart of condensed matter physics. They are naturally described in terms of emergent quasiparticles that encode the internal dynamics of matter, such as phonons in crystals \cite{AshcroftMermin}, plasmons in electron liquids \cite{PinesNozieres}, or topological quasiparticles in Dirac and Weyl materials \cite{Armitage2018}. Understanding how quasiparticles respond to crystalline defects is therefore a fundamental problem, since defects represent topological singularities of the lattice that cannot be removed by smooth deformations and can strongly reshape spectra, induce bound states, and modify transport \cite{friedel1964dislocations,Teo2010,Slager2014,Lund2019,Churochkin2022,Fall2025}. In ordered magnets, magnons, the quanta of spin-wave excitations, govern spin and thermal transport and provide a paradigmatic platform for bosonic band topology \cite{Chumak2015,Mook2014}. While crystalline defects are usually regarded as static perturbations, their quantized dynamics can itself give rise to emergent collective excitations.

Dislocations--line defects characterized by a discontinuity in the lattice--introduce an intrinsically topological background through their Burgers vector and associated singularity \cite{friedel1964dislocations,HirthLothe1982}. In electronic, photonic, and acoustic systems, they can host robust defect-bound states and one-dimensional modes protected by lattice topology \cite{Ran2009,Teo2010,Slager2014,Xue2021,Ye2022}. Similar mechanisms are expected to reshape magnon spectra and generate localized spin-wave modes, even in the absence of disorder \cite{SajiPRB2025,saji2025-arxiv,larronde2026scattering}. These results establish crystalline defects as fundamental elements of quasiparticle topology rather than perturbative imperfections.

The quantized dislocation vibrations--dislons--provide a microscopic framework for describing dislocation dynamics within many-body theory \cite{li2017tailoring,li2018theory}. While dislon interactions with electrons and phonons have been explored \cite{li2018theory}, their coupling to magnetic excitations remains largely unexplored, despite early indications of defect-induced magnetoelastic effects on spin dynamics \cite{Fomethe1982}. More fundamentally, dislocation motion obeys topological constraints, such as Burgers-vector conservation and glide--climb asymmetry, which can be naturally captured within the fracton--elasticity duality, where dislocations correspond to dipolar charges of higher-rank gauge theories \cite{PretkoPRL2018,PretkoPRB2019}.
\begin{figure}[tbh]
\includegraphics[width=0.8\columnwidth]{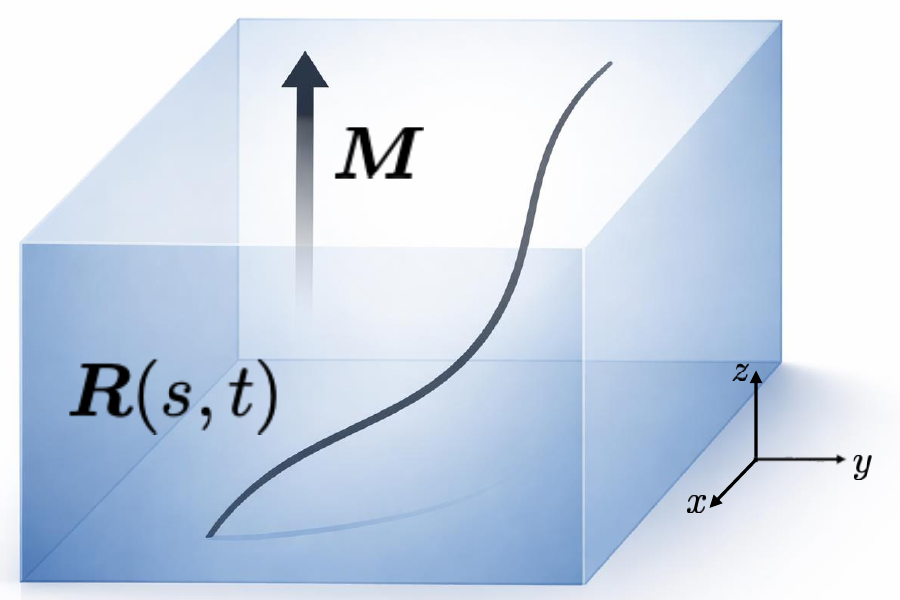}
\caption{Schematic of a dynamical dislocation in a three-dimensional magnetic crystal. Transverse fluctuations of the dislocation line $R^{\mu}(s,t)$ define dislon excitations that hybridize with magnons propagating around the equilibrium magnetization ${\bs M}$.}
\label{fig:dislons-setup}
\end{figure}

In this Letter, we develop a microscopic theory for the coupled dynamics of magnons, dislons, and phonons in magnetically ordered crystals. Using fracton--elasticity duality, we show that dislocation dynamics coherently hybridizes with magnons, giving rise to magnon--dislon collective modes. We thus identify dislons as fundamental dynamical degrees of freedom in magnetic solids, placing defect-mediated coupling on the same footing as conventional hybrid magnon systems. While hybrid magnon platforms involving photons, phonons, and other magnons are well established \cite{Huebl2013,Tabuchi2015,Zhang2014,Demokritov2001,streib2019magnon,SimensenPRB2019,SimensenPRB2020,MellaPRB2024}, coupling to dislons has not been explored. Our work identifies distinct signatures of dynamical topological defects, including helicity-selective hybridization, dark dislon branches, and glide-induced ellipticity, thereby establishing magnon-dislon hybridization as a distinct class of hybrid bosonic excitation.

\textit{Magnetic and elastic theory.--} We consider a ferromagnetic crystal containing embedded dislocation defects, see Fig. \ref{fig:dislons-setup}. The dislocation is described by the space-time dependent curve $R^{\mu}(s,t)$, where $s$ parametrizes the defect line, while $\mu=0$ and $\mu=1,2,3$ denote the temporal and spatial coordinates, respectively. Fluctuations of this curve describe transverse vibrations which define the dislon field and enable their coupling to magnons. Each dislocation carries a Burgers vector ${\bs b}$ that characterizes the lattice distortion and acts as a conserved topological charge \cite{HirthLothe1982}. Here we derive an effective theory for vibrations of dislocations and the magnetization via fracton–elasticity duality.

We begin describing the magnetic and elastic degrees of freedom by the total action ${\cal S}_T = {\cal S}_{\text{el-}m} + {\cal S}_{m}$, in a 3D crystal with volume $V=L_xL_yL_z$. The first term describes the elastic and magnetoelastic contributions,
\begin{align*}
{\cal S}_{\text{el-}m}=\frac{1}{2}\int d{\bs x}dt\left[\rho_{\text{el}}(\partial_t {\bs u})^2- (C^{ijkl}\varepsilon_{ij}+2\Sigma^{kl})\varepsilon_{kl}\right],
\end{align*}
where $\rho_{\text{el}}$ is the mass density of the elastic medium, ${\bs u}({\bs x},t)$ the lattice displacement field, and $\varepsilon_{ij} = (\partial_i u_j + \partial_j u_i)/2$ the symmetric strain tensor, while $C^{ijkl}$ is the fourth-rank elastic tensor determined by the lattice symmetries. The magnetoelastic coupling between strain and magnetization field ${\bs m}$, satisfying $|{\bs m}|=1$, enters through $\Sigma^{kl}[{\bs m}]=B^{ijkl}m_i m_j$, with $B^{ijkl}$ the magnetoelastic tensor. The magnetization dynamics is described by the action
${\cal S}_{m}=\int d{\bs x}dt\,\left({\cal K}[{\bs m}]-{\cal E}[{\bs m}]\right)$, where the first term represents the Berry-phase kinetic term. In the
micromagnetic approximation the magnetic energy density is  $\mathcal{E}[{\bs m}]=A(\nabla {\bs m})^2-K m_z^2-\mu_{0} M_{s} H m_{z}$, where $A$ is the exchange stiffness, $K>0$ the easy-axis anisotropy, and Zeeman coupling with an external magnetic field $H$ applied along $z-$direction, with $M_{s}$ the saturation magnetization and $\mu_{0}$ is the vacuum permeability.

In crystalline solids with topological defects, elasticity admits an equivalent formulation in terms of gauge fields, establishing a direct link to fracton-like behavior \cite{PretkoPRL2018,PretkoPRB2019}. It is useful to decompose the strain tensor into regular and singular parts, $\varepsilon_{ij}=\bar{\epsilon}_{ij}+\big(\partial_i \tilde u_j+\partial_j \tilde u_i\big)/2$, where $\tilde u_i$ is smooth and single-valued, whereas $\bar{\epsilon}_{ij}$ contains the nontrivial topology associated with defects \footnote{It satisfies a compatibility relation of the form $\varepsilon^{ik}\bar{\epsilon}^{j\ell}\partial_i\partial_j \bar{\epsilon}_{k\ell}=\rho$, with $\rho$ encoding the defect density.}. This decomposition is ambiguous: shifts of $\bar{\epsilon}_{ij}$ by regular gradients leave physical configurations invariant, giving rise to an emergent gauge structure. To make this explicit, we introduce auxiliary Hubbard--Stratonovich fields, the momentum density $\pi^i$ and symmetric stress tensor $\sigma^{ij}$, see Supplemental Material (SM)-\ref{sec:totalaction} for details. Defining the stress--momentum tensor $T^{0I}=\pi^I$ and $T^{iI}=\sigma^{iI}+\Sigma^{iI}$, with $\Sigma^{iI}$ the defect contribution, the equations of motion impose the continuity equation,
\begin{equation}\label{eq:stress-momentum}
\partial_\mu T^{\mu I}=0,
\end{equation}
expressing local momentum conservation in the presence of defects \footnote{This constraint can be solved identically by $T=* d\mathcal{A}$, where $*$ denotes the Hodge-star operator and ${\cal A}$ the gauge field.}. Eq.~(\ref{eq:stress-momentum}) is solved exactly by an antisymmetric rank-2 gauge field $\mathcal{A}^{I}_{\mu\nu}$, such that $T^{\mu I}=\epsilon^{\mu\nu\lambda\rho}\partial_\nu \mathcal{A}^{I}_{\lambda\rho}$, which corresponds to a generalized Bianchi identity. The freedom in representing $T^{\mu I}$ implies the gauge invariance $\mathcal{A}^{I}_{\mu\nu}\rightarrow \mathcal{A}^{I}_{\mu\nu}+\partial_\mu \xi^I_\nu-\partial_\nu \xi^I_\mu$, where $\xi^I_\mu$ is a smooth and single-valued vector field. Within this dual framework, elastic modes appear as gauge fluctuations and defects as their sources.

Replacing the components of the stress--momentum tensor into the action, ${\cal S}_{\mathrm{el}-m}$, the theory can be recast as $\mathcal{S}_{T}= \tilde{\mathcal{S}}_{\mathcal{A}}+\mathcal{S}_{\mathcal{A}-m}+{\cal S}_m$. The first contribution, $\tilde{\mathcal S}_{\mathcal{A}}=\mathcal{S}_{\mathcal{A}}+S_{\mathrm{KR}}$, defines the gauge sector associated with the dual elastic description \cite{PretkoPRL2018,PretkoPRB2019}. The free gauge modes are described, in Fourier space, by 
\begin{align}
\mathcal{S}_{\mathcal{A}}= \int (\phi_{\alpha i}^{I})^{T}(-q)\,[G^{IJ}_{\alpha\beta,ij}]^{-1}(q)\,\phi_{\beta j}^{J}(q)\,d^4q,
\end{align}
where we have introduced the fields $\phi_{Bi}^{I}=\epsilon_{ijk}\mathcal{A}_{jk}^{I}$ and $\phi_{Ai}^{I}=\mathcal{A}_{0i}^{I}$, with $\alpha,\beta=A,B$. The gauge-field propagator ${G}^{IJ}_{\alpha\beta,ij}(q)$, detailed in SM-\ref{sec:propagator}, encodes the elastic (fractonic) response of the medium. Topological defects couple minimally to the gauge field through the Kalb--Ramond term \cite{Lin2025}, 
\begin{align}
S_{\mathrm{KR}}=-\frac{1}{2}\int d{\bs x}dt\,(\mathcal{J}_{\mathrm{d}})^{\mu\nu}_{I}({\bs x},t)\,\mathcal{A}^{I}_{\mu\nu}({\bs x},t),
\end{align}
where the dislocation current $(J_{\mathrm d})^{\mu\nu}_{I}=\epsilon^{\lambda\rho\mu\nu}\partial_{\lambda}\partial_{\rho}u^{I}_{\mathrm d}$ captures the singular part of the displacement field. Equivalently, it can be represented as the current of a dislocation line, {$(J_{\mathrm d})^{\mu\nu}_{I}(x)=b_I\int ds\,dt'\,\epsilon^{ab}\partial_aR^\mu\partial_bR^\nu\delta^{(4)}(x-R(t',s))$}. This term shows that dislocations act as gauge-field sources, analogous to charges, and encodes their constrained mobility within the fracton framework \cite{PretkoPRB2019}. The coupling with the magnetization reads,
\begin{equation}
\mathcal{S}_{\mathcal{A}-m}=-\frac{1}{2}\int d{\bs x}dt\,(\mathcal{J}_{m})^{\nu\lambda}_{I}({\bs x},t)\,\mathcal{A}^{I}_{\nu\lambda}({\bs x},t),
\end{equation}
with the current $(\mathcal{J}_{m})^{\nu\lambda}_{I}=-2C^{-1}_{Ijkl}\,\epsilon^{j\mu\nu\lambda}\,\partial_{\mu}\Sigma^{kl}[{\bm m}]$, behaving as a source that couples the spin to the dual elastic sector. 
Altogether, the action $\mathcal S_T$ unifies defects, elasticity, and spin dynamics within a rank-2 gauge theory.

Integrating out the gauge fields in the partition function $\mathcal{Z}=\int \mathscr{D}[{\bm m}]\mathscr{D}[{R}]\mathscr{D}[\mathcal{A}]e^{i\mathcal{S}_{T}[{\bs m}, {R},{\cal A}]/{\hbar}}$, yields an effective theory for the coupled magnetization and dislocation dynamics. Since the action is quadratic in $\mathcal{A}$, the integration is exact, yielding $\mathcal{S}_{\mathrm{eff}}[{\bs m}, {R}] ={\mathcal{S}}_{m} + \mathcal{S}_{R} + \mathcal{S}_{c}$. Here, $\mathcal{S}_{R}=T_{0}\int dtds\,\eta^{ab}\partial_{a}R^{\mu}\partial_{b}R_{\mu}$ is the Polyakov action governing dislocation dynamics, with $T_{0}$ the string tension. The coupling takes the current–current form
\begin{equation}\label{eq:DeltaS}
\mathcal{S}_{c}=\int ({\mathcal{J}}_{\alpha})^{i}_{I}(-q)\,{G}^{IJ}_{\alpha\beta,ij}(q)\,({\mathcal{J}}_{\beta})^{j}_{J}(q)\, d^4q,
\end{equation}
where the total current ${\mathcal{J}}_{\alpha} = {\mathcal{J}}_{m,\alpha} + {\mathcal{J}}_{d,\alpha}$ combines magnetic and dislocation contributions, with $({\mathcal{J}}_{A})^{i}_{I} = \mathcal{J}^{i0}_{I}$ and $({\mathcal{J}}_{B})^{i}_{I} = \epsilon^{ijk}\mathcal{J}^{jk}_{I}/2$ corresponding to the temporal and spatial sectors, respectively. Because ${\mathcal{J}}_{d,\alpha}$ carries the Burgers vector, Eq.~(\ref{eq:DeltaS}) shows that magnon--dislon coupling encodes defect topology and mobility constraints beyond conventional magnon--phonon physics. For an isotropic elastic solid $C_{ijkl}=\lambda\delta_{ij}\delta_{kl}+\mu(\delta_{ik}\delta_{jl}+\delta_{il}\delta_{jk})$ (and similarly for $B_{ijkl}$), where $\lambda$ and $\mu$ are the Lamé coefficients, the magnetoelastic current reads
\begin{align}
(\mathcal{J}_{m,A})^{i}_{I} &\label{eq:JmA}= 2B_{2}\,\mu^{-1}\,\epsilon^{ijk}(m_{j}\partial_{k}m_{I}+m_{I}\partial_{k}m_{j}),\\
(\mathcal{J}_{m,B})^{i}_{I} &\label{eq:JmB}= 2B_{2}\,\mu^{-1}\,(m_{i}\partial_{t}m_{I}+m_{I}\partial_{t}m_{i}),
\end{align}
 For a straight dislocation, the world sheet is parametrized as
$R^{\mu}=R_{\parallel}^{\mu}+R_{\perp}^{\mu}$,
with $R_{\parallel}^{\mu}=(t,0,0,z)$ and
$R_{\perp}^{\mu}(t,z)=(0,X_1(t,z),X_2(t,z),0)$, where $X_1$ and $X_2$ are the transverse dislon fields. Thus the dislocation current is
\begin{align}
(\mathcal{J}_{\mathrm{d},A})_{I}^{i} &\label{eq:JdA}=b_{I}(\delta^{i}_{z}+ \partial_{z}R_{\perp}^{i}-\delta^{i}_{z} R_{\perp}^{j}\partial_{j}) \delta^{2}({\bs x}^{\perp}), \\
(\mathcal{J}_{\mathrm{d},B})_{I}^{i} &\label{eq:JdB}= -b_{I}\epsilon_{ij}\partial_{t}R_{\perp}^{j} \delta^{2}({\bs x}^{\perp}),
\end{align}
up to linear order in ${R}^{\mu}_{\perp}$ and ${\bs x}_{\perp}$ the coordinates in the transverse $xy-$plane.

Substituting the total current into Eq.~(\ref{eq:DeltaS}) yields magnetic ($\mathcal S_c^{mm}$), dislocation ($\mathcal S_c^{RR}$), and mixed ($\mathcal S_c^{mR}$) interactions mediated by the non-local elastic propagator, see SM-\ref{sec:action-coupling}. In the absence of magnetic degrees of freedom, one recovers the effective theory of dislocations coupled to bulk phonons derived in Ref.~\cite{Lin2025}. More importantly, dislocations qualitatively modify the magnetic ground state. In the static limit, $\mathcal{J}_{B}=0$ [cf. Eqs.~(\ref{eq:JmB}) and (\ref{eq:JdB})], the mixed contribution reduces to a Lifshitz-type invariant $\mathcal S^{mR}_{c,\mathrm{stat}}=\int d^3r\,D^{Jlk}(\mathbf r_\perp)\left(m_l\partial_k m_J+m_J\partial_k m_l\right)$, with
\begin{equation}
D^{Jlk}(\mathbf r_\perp)=\frac{4B_2}{\mu}
\,b_I\epsilon^{jlk}\int dz'\,G^{IJ}_{AA,zj}(\mathbf r_\perp,-z').
\end{equation}
The coupling $D_{Jlk}$ is local and determined by the dislocation geometry, promoting chiral magnetic textures and localized mode bound to the defect core \cite{larronde2026scattering,Latorre2026}. Beyond the static limit, the retarded propagator dynamically couples magnons and dislons through the underlying fractonic response.

{\it Magnon--dislon quasiparticles.--} We now consider small magnetic fluctuations coupled to dislocation vibrations. The magnetic state is assumed to be saturated along the $z-$axis, with spin-wave fluctuations confined to the transverse $xy-$plane. Since the dislocation dynamics are localized near the line core, we focus on quasi-one-dimensional magnon modes propagating along the dislocation axis. Accordingly, the magnonic fluctuations are expanded as $\delta m_i=\sum_{nm}\psi_{i}^{nm}({\bs q})\delta\left(q_x-2\pi nL^{-1}_x\right)\delta\left(q_y-2\pi mL^{-1}_y\right)$, with $n,m\in \mathbb{Z}$. The transverse momenta are thus quantized, yielding a discrete set of magnon channels in the regime $L_x,L_y\ll L_z$. Focusing on the fundamental mode $n=m=0$, i.e., the ${\bs q}_{\perp}=0$ homogeneous transverse mode, we define $\delta m_{\pm}=\delta m_{x}\pm i\delta m_{y}$, and analogously $X_{\pm}=X_{1}\pm i X_{2}$, The coupled magnon--dislon system is described by the effective action
\begin{equation}
\label{eq:effective-action-MD}
\mathcal{S}_{\mathrm{eff}}
=
\int d\kappa \;
\Psi_{\kappa}^{\dagger}
\Lambda_{\kappa}
\Psi_{\kappa},
\qquad
\Lambda_{\kappa}=
\begin{pmatrix}
\Lambda_{m,\kappa} & \Xi_{\kappa}\\
\Xi^{\dagger}_{\kappa} & \Lambda_{d,\kappa}
\end{pmatrix},
\end{equation}
where $\Psi_{\kappa}=(\delta m_{+},\delta m_{-},X_{+},X_{-})^{T}$ and $\kappa=(\omega,q_z)$ denotes frequency of the hybridized eigenmode and momentum along the dislocation line, see SM-\ref{sec:magnon-dislons} for details. The magnon sector is governed by the matrix $\Lambda_{m,\kappa}
=\omega\sigma_{z}+\zeta_{\kappa}\sigma_{x}-\omega_{m}(q_z)\sigma_{0}$, with $\omega_{m}(q_z)=\omega_{0}+{2A\gamma}q_z^{2}/{M_{s}}$ the bare magnon dispersion and $\omega_{0}=\gamma
\left(\mu_{0}H+{2K}/{M_{s}}\right)$ the gap. The coefficient $\zeta_{\kappa}=8\mu^{-2}
{(B_{2}c_{T}\omega)^{2}}/(
{c_{T}^{2}q_z^{2}-\omega^{2}})$ corresponds to the magnetoelastic self-energy induced by the elastic gauge field, with $c_T=\sqrt{{\mu}/{\rho_{\rm el}}}$ and $c_L=\sqrt{({\lambda+2\mu})/{\rho_{\rm el}}}$ the transverse and longitudinal sound velocity, respectively. Here ${\boldsymbol \sigma}$ denotes the vector of Pauli matrices and $\sigma_{0}$ the identity matrix. 

For screw dislocations, with Burgers vector $\bs b_s=b\hat{\bs z}$, rotational symmetry around the defect axis is preserved. In the circular basis, the dislon sector is therefore diagonal, $\Lambda^{s}_{d,\kappa}=(\rho(q_z)\omega^2-T_{s}(q_z)q_z^2)\sigma_0$, yielding two degenerate circularly polarized modes, $X_\pm$, with dispersion $\omega_{d,s}(q_z)=|q_z|\sqrt{T_s(q_z)/\rho(q_z)}$. The effective mass density, $\rho(q_z)=\rho_0-\rho_{\rm el}b^2\ln(q_z^2L^2)/(4\pi)$, and line tension, $T_{s}(q_z)=T_0+\mu b^2(1-c_T^2/c_L^2)\ln(q_z^2L^2)/\pi$, acquire logarithmic renormalizations from the elastic gauge-field environment \cite{Lin2025}, where $L$ is an infrared cutoff set by the dislocation length. In the bare elastic-string limit the dislon velocity reduces to $v_d=\sqrt{T_0/\rho_0}=c_T$, since $T_0\simeq {\mu b^2}\ln\!\left({L}/{a}\right)/{4\pi}$ and $\rho_0\simeq \rho_{\rm el} b^2\ln\!\left({L}/{a}\right)/{4\pi}$ with $a$ the core cutoff. The magnon--dislon interaction is encoded in the off-diagonal block $\Xi^{s}_{\kappa}=\Gamma^{s}_{\kappa}\sigma_y$, where $\Gamma^{s}_{\kappa}=(b\mu/2B_2)\zeta_{\kappa}$ is the magnetoelastic vertex.
\begin{figure}[tbh]
\includegraphics[width=8.6cm]{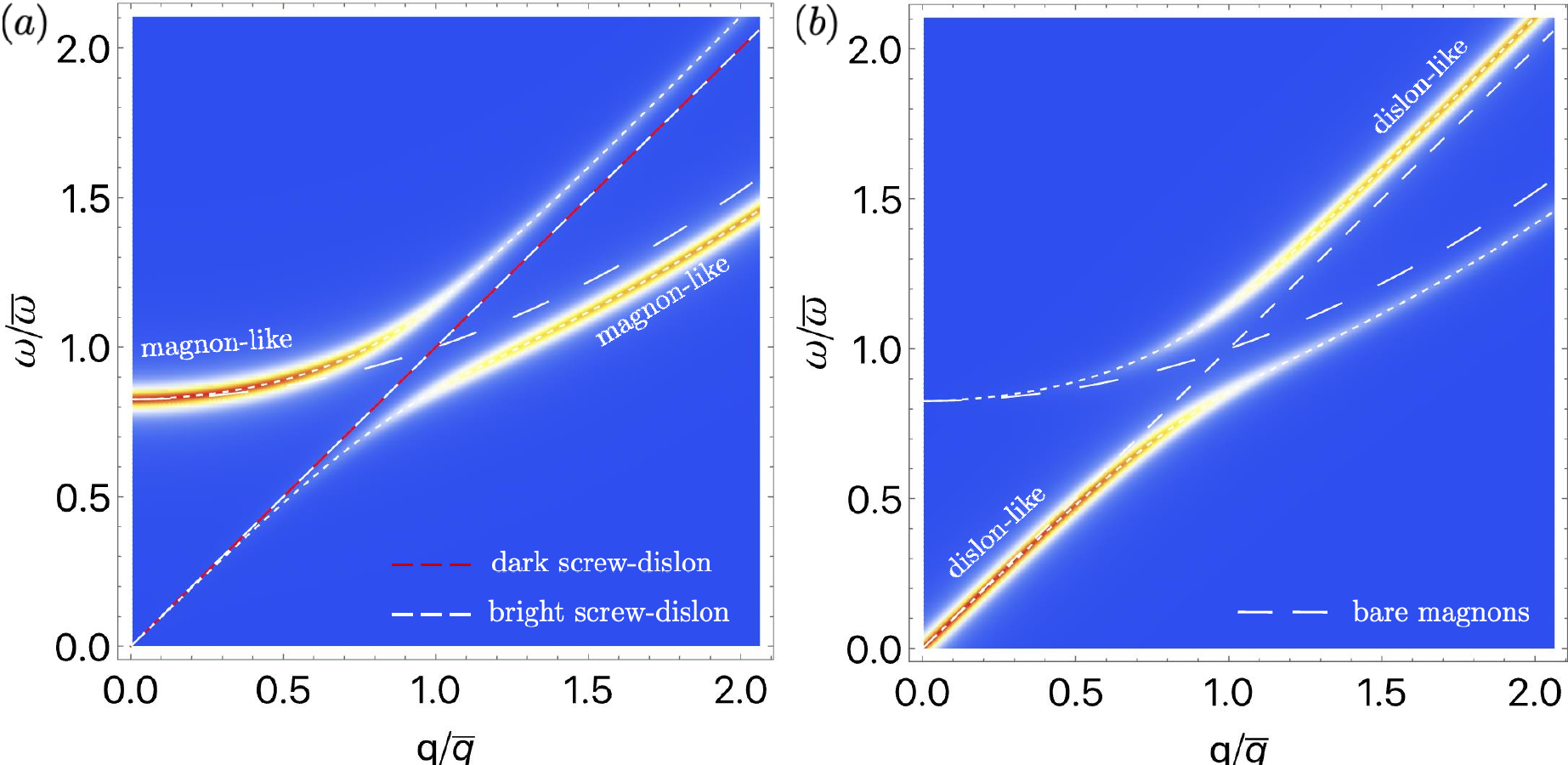}
\caption{Hybridization between magnons and screw-dislons. (a) Magnon and (b) dislon spectral weights. The avoided crossing signals coherent hybridization and spectral-weight transfer. Far from resonance, the hybrid modes recover their magnon- or dislon-like character. The red dashed line in (a) marks the symmetry-protected dark screw-dislon branch, which remains uncoupled to magnons. White dotted and dashed lines denote the hybridized and bare dispersions, respectively.}
\label{fig:magnon-dislon}
\end{figure}

The hybridization between magnons and screw dislons is governed by the poles of the kernel, $\det[\Lambda^{s}_{\kappa}]=0$.  The magnon-dislon coupling is controlled by the resonant vertex $\Gamma^{{s}}_{\kappa}\propto \omega^{2}[c_T^{2}q_z^{2}-(\omega+i\eta)^2]^{-1}$, whose singular structure originates from the transverse elastic propagator. Here, $\eta$ regularizes the transverse phonon pole and phenomenologically accounts for damping and finite elastic lifetimes. Since the bare screw-dislon dispersion satisfies $\omega_{d,s}(q_z)\simeq c_T|q_z|$, the hybridization is resonantly enhanced near the crossing points defined by $\omega_m(\bar q)=\omega_{d,s}(\bar q)\equiv\bar\omega$. Real crossings exist only when $\Delta_T=c_T^2-4J\omega_0>0$, yielding two resonant momenta $\bar q$. In the weak-coupling regime, the coupled modes exhibit an avoided crossing with hybrid branches $\omega_\pm(q_z)=\omega_+(q_z)\pm\sqrt{\omega_-^2(q_z)+g^2}$, where $\omega_{\pm}(q_z)=[\omega_m(q_z)\pm\omega_{d,s}(q_z)]/2$ and $g={(B_2bc_T^2\sqrt{2\bar{\omega}})}/{(\mu\eta\sqrt{\rho_0})}$. The corresponding hybridization gap is $\Delta_{\mathrm h}=2g$, and the perturbative description remains valid for $g\ll\bar\omega$. In this resonant regime the hybrid modes acquire a mixed magnon--dislon character and finite linewidths controlled by $\eta$. Fig.~\ref{fig:magnon-dislon}(a) and (b) show the magnon and dislon spectral functions, revealing the avoided crossing and the associated transfer of spectral weight. Away from resonance, the hybrid branches recover the magnon-like or dislon-like character.  The results are presented in dimensionless units normalized by the bare magnon--dislon resonance, with ${\omega_0}=0.8{\bar\omega}$, ${J\bar q^2}=0.2{\bar\omega}$, and ${\eta}=0.03{\bar\omega}$. These parameters are representative of a weakly damped gapped magnon branch coupled to a dislon mode in clean insulating ferro- and ferrimagnets \cite{GurevichMelkov1996}.

For screw dislocations, rotational symmetry about the defect axis allows the excitations to be classified by their helicity, defined as the projection of angular momentum along the dislocation line. The circularly polarized dislon modes ($X_\pm$) correspond to opposite senses of transverse rotation of the dislocation core and therefore carry opposite helicities. 
As shown in SM-\ref{sec:screw-dislons}, the interaction $\Xi^{s}_{\kappa}$ is helicity selective: $\delta m_{+}$ couples only to $X_-$, whereas $X_+$ remains uncoupled. Consequently, only one dislon helicity hybridizes with magnons, while the opposite-helicity branch remains a symmetry-protected dark mode [Fig.~\ref{fig:magnon-dislon}(a)]. This selection rule follows from angular-momentum conservation about the dislocation axis and ensures that the hybrid excitations retain a well-defined helicity. 

\begin{figure}[tbh]
\includegraphics[width=\columnwidth]{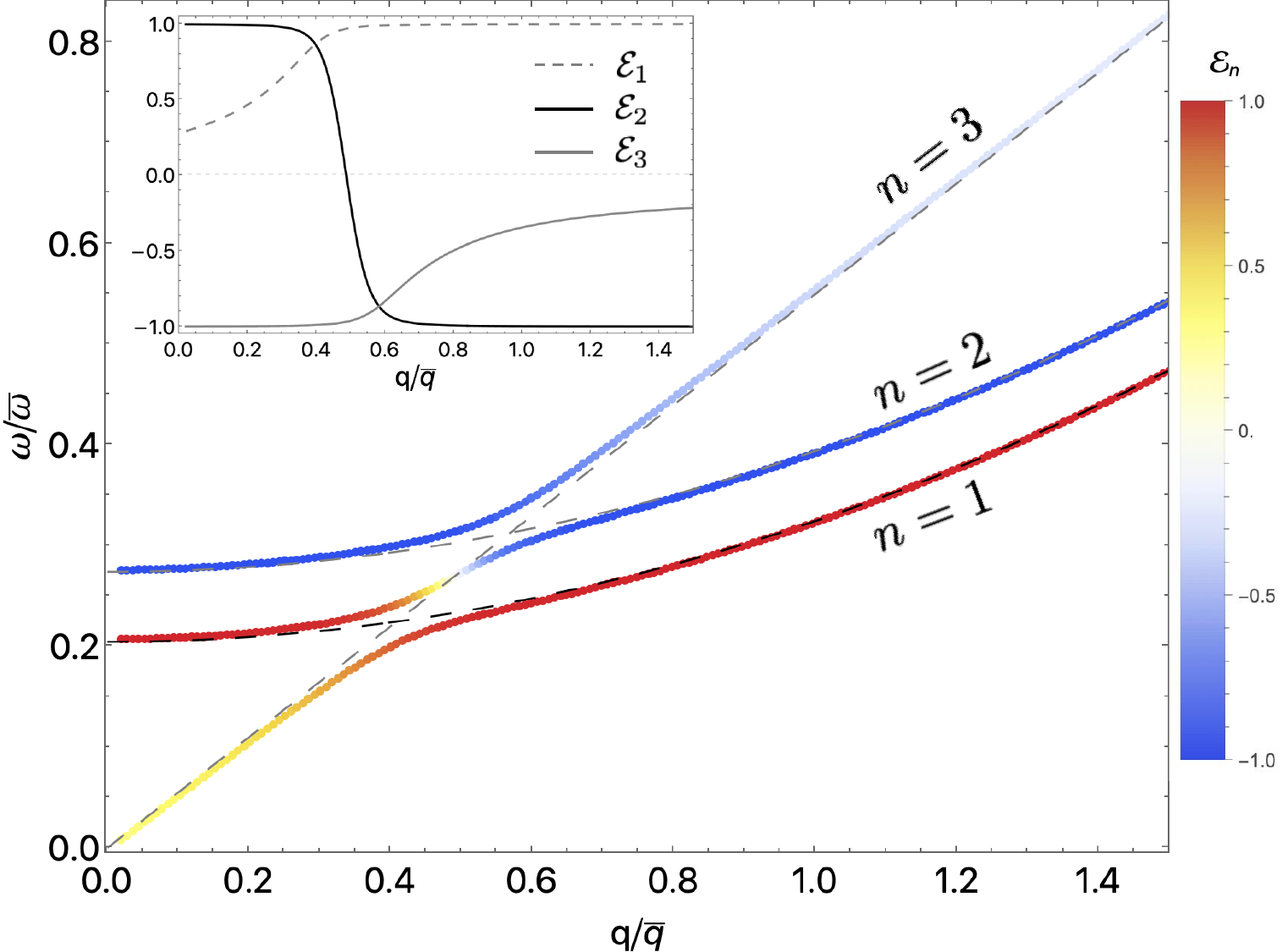}
\caption{Hybrid edge magnon–dislon spectrum, with the branches $\omega_n(q_z)$, colored according to their ellipticity $\mathcal E_n$, for $\omega_0=0.8\bar\omega$, $J\bar q^2=0.2\bar\omega$, $g_{\text{eff}}=0.15\bar\omega$, with $g_{\text{eff}}=\sqrt{g^2_{10}+g^2_{01}}$, and $\eta=0.03\bar\omega$. The avoided crossing at $q_z/\bar q \simeq 0.5$ reflects resonant magnon–dislon hybridization and the accompanying exchange of magnonic and dislonic character. Inset: Ellipticity of the three hybrid branches and its redistribution near resonance.}
\label{fig:magnon-edge-dislon}
\end{figure}

For edge dislocations, the Burgers vector $\bs b_e=b\hat{\bs x}$ breaks rotational symmetry in the transverse plane. Consequently, the circular basis no longer diagonalizes the dislon sector, leading to anisotropic hybridization. At low energies, the glide constraint suppresses climb fluctuations, leaving a single propagating dislon mode \(X_g\) polarized along \(\bs b_e\). Unlike the screw case, the magnon--dislon vertex vanishes in the limit \({\bs q}_\perp=0\), so the homogeneous magnon mode $\psi^{00}_i$ remains decoupled. Hybridization is restored by magnon states carrying finite transverse momentum, with the leading contributions arising from the first transverse harmonics $\psi^{10}_i$ and $\psi^{01}_i$, corresponding to momenta \(\bs q_{10}=(2\pi/L_x,0,q_z)\) and \(\bs q_{01}=(0,2\pi/L_y,q_z)\). The resonant sector is therefore spanned by \(\Phi=(\psi^{10},\psi^{01},X_g)^T\), and the hybrid spectrum follows from the effective kernel
\begin{align*}
\Lambda_{\rm eff}(q_z)=
\begin{pmatrix}
\omega_{10}(q_z)-\omega & 0 & g_{10}
\\
0 & \omega_{01}(q_z)-\omega & g_{01}
\\
g_{10}^{*} & g_{01}^{*} & \omega_{d,e}(q_z)-\omega
\end{pmatrix},
\end{align*}
where the coupling $g_{10}=|\Gamma_{+g}^{10}|$ and $g_{01}=|\Gamma_{+g}^{01}|$, given at the SM-\ref{sec:edge-dislons}, inherit the momentum and frequency dependence of the vertex. Near the avoided crossing, however, they are evaluated at the bare resonance and treated as effective constants. The spectrum consists of three hybrid modes arising from the coupling between the glide dislon and the two lowest transverse magnon harmonics. For the generic case \(L_x\neq L_y\), the modes \(\psi^{10}\) and \(\psi^{01}\) are nondegenerate, yielding the three hybrid branches shown in Fig.~\ref{fig:magnon-edge-dislon}, where the color scale denotes the ellipticity.

In contrast to screw dislons, edge dislons are polarized predominantly along the Burgers-vector direction as a consequence of the glide constraint. The hybridization transfers this anisotropy to the magnon sector and deforms the spin precession from circular to elliptical. We quantify this effect through the ellipticity $\mathcal E_n(q_z)=
{(|\delta m_x^{(n)}|^2-|\delta m_y^{(n)}|^2})/(
{|\delta m_x^{(n)}|^2+|\delta m_y^{(n)}|^2}),$
where $\delta m_x^{(n)}$ and $\delta m_y^{(n)}$ are the transverse spin amplitudes of the $n$th hybrid mode (see SM~\ref{sec:edge-dislons}). As shown in Fig.~\ref{fig:magnon-edge-dislon}, the hybrid modes acquire finite ellipticity, reflecting the anisotropic coupling to the glide-polarized edge dislon. Near the avoided crossing, where magnon and dislon components are maximally mixed, the ellipticity is strongly redistributed among the three branches (inset), accompanying the exchange of magnonic and dislonic character. The polarization structure provides a direct signature of edge-dislocation hybridization.

The predicted hybrid spectrum should be accessible through momentum-resolved probes such as inelastic neutron scattering~\cite{Lovesey1984} and Brillouin light scattering~\cite{Demokritov2001}. While the avoided crossing is a generic feature of hybrid bosonic excitations, magnon--dislon hybridization is distinguished by polarization signatures tied to dislocation geometry. In particular, screw dislocations support helicity-selective hybridization and a symmetry-protected dark dislon branch, whereas edge dislocations generate finite spin-precession ellipticity through the glide constraint. These defect-specific signatures provide a direct means to distinguish magnon--dislon hybridization from conventional bulk magnon--phonon coupling. In addition, local probes such as NV magnetometry~\cite{Rondin2014,Casola2018} and vibrational STEM-EELS~\cite{Hage2022,Liu2024} could identify the defect origin of the hybrid modes through their localization near individual dislocation cores.

{\it Conclusions.--} We have developed a microscopic theory of magnon--dislon hybridization in magnetic insulators. Within the fracton--elasticity dual framework, dislocations emerge as dynamical gauge sources whose Burgers-vector current imposes geometry-dependent selection rules on spin excitations. Screw dislocations preserve axial angular momentum, producing helicity-selective hybridization and symmetry-protected dark dislon modes, whereas edge dislocations break cylindrical symmetry through the glide constraint and generate finite spin-precession ellipticity. These geometry-dependent polarization fingerprints distinguish magnon--dislon hybridization from conventional bulk magnon--phonon hybridization. More broadly, our results establish dynamical topological defects as a new platform for hybrid bosonic excitations in magnetic materials.

{\it Acknowledgements} R.E.T. and N.V.S. acknowledge funding from Fondecyt Regular 1230747 and 1250364, respectively.

\bibliography{dislons}

\onecolumngrid
\section{Supplemental Material}
{\subsection{Derivation of the total action}\label{sec:totalaction}
The gauge structure of elasticity in the presence of topological defects, is made evident by introducing a Hubbard--Stratonovich transformation and rewrite the elastic action in terms of the momentum density \(\pi^i\) and the symmetric stress tensor \(\sigma^{ij}\). The elastic Lagrangian becomes
\begin{equation}
\mathcal{L}_{\mathrm{el}}=\frac{1}{2\rho_{\mathrm{el}}}\pi_i\pi^i+\frac{1}{2}C^{-1}_{ijkl}\sigma^{ij}\sigma^{kl}
+\pi^i\partial_tu_i-\sigma^{ij}\partial_i u_j,
\end{equation}
where \(C^{-1}_{ijkl}\) denotes the elastic compliance tensor. The displacement field \(u_i\) appears linearly and therefore acts as a Lagrange multiplier. Integrating over \(u_i\) imposes the local conservation law $\partial_t\pi^i-\partial_j\sigma^{ji}=0$,
which expresses momentum conservation in the elastic medium. Including the magnetoelastic contribution and introducing the stress--momentum tensor $T^{0I}\equiv \pi^I$ and $T^{iI}\equiv \sigma^{iI}+\Sigma^{iI}$, the elastic--magnetic action can be written as
\begin{align}
\mathcal{S}_{\mathrm{el-m}}&=\frac12
\int d^3x\,dt\left[\frac{1}{\rho_{\mathrm{el}}}T^{0I}T^{0I}+C^{-1}_{IJAB}\bigl(T^{IJ}-\Sigma^{IJ}\bigr)\bigl(T^{AB}-\Sigma^{AB}\bigr)\right].
\label{eq:Selm_T}
\end{align}
The conservation law is solved identically by introducing an antisymmetric rank-two gauge field \(A^I_{\mu\nu}=-A^I_{\nu\mu}\), $T^{\mu I}=\epsilon^{\mu\nu\alpha\beta}\partial_\nu A^I_{\alpha\beta}$, which is substituted into Eq.~(\ref{eq:Selm_T}), yielding
\begin{align}
\mathcal{S}_{\mathrm{el-m}}=\int d^3x\,dt
\left[\mathcal{L}_0[A]+C^{-1}_{ijab}\epsilon^{j\nu\alpha\beta}\partial_\nu A^i_{\alpha\beta}\Sigma^{ab}
\right],
\end{align}
which naturally decomposes as $\mathcal{S}_{\mathrm{el-m}}=
\mathcal{S}_{A}+\mathcal{S}_{\mathrm{KR}}+\mathcal{S}_{A-m}+\Delta\mathcal{S}$. Here \(\mathcal{S}_{A}\) describes the free elastic gauge sector, \(\mathcal{S}_{\mathrm{KR}}\) is the Kalb--Ramond coupling between the gauge field and dislocations, \(\mathcal{S}_{A-m}\) represents the coupling between the gauge field and magnetization, and
\begin{equation}
\Delta\mathcal{S}=\frac12 \int d^3x\,dt\,\Sigma^{ij}C^{-1}_{ijkl}\Sigma^{kl}
\end{equation}
is the magnetoelastic correction generated by integrating out the stress degrees of freedom. For a three-dimensional isotropic solid, $C_{ijkl}=\lambda\,\delta_{ij}\delta_{kl}+\mu\left(\delta_{ik}\delta_{jl}+\delta_{il}\delta_{jk}\right)$, where \(\lambda\) and \(\mu\) are the Lamé coefficients. Similarly, the isotropic magnetoelastic tensor is $B_{ijkl}=B_1\delta_{ij}\delta_{kl}+B_2
\left(\delta_{ik}\delta_{jl}+\delta_{il}\delta_{jk}\right)$. Using \(\Sigma^{ij}=B_{abij}m_am_b\), one obtains
\begin{equation}
\Sigma^{ij}C^{-1}_{ijkl}
\Sigma^{kl}=B_{abij}C^{-1}_{ijkl}B_{cdkl}
\,m_am_bm_cm_d,
\end{equation}
with
\begin{align}
B_{abij}C^{-1}_{ijkl}B_{cdkl}=\frac{
3B_1^{\,2}\mu+4B_1B_2\mu-2B_2^{\,2}\lambda
}{\mu(3\lambda+2\mu)}\delta_{ab}\delta_{cd}
+\frac{B_2^{\,2}}{\mu}\left(\delta_{ac}\delta_{bd}+\delta_{ad}\delta_{bc}\right).
\end{align}
Since the magnetization is normalized, ${\bs m}\cdot{\bs m}=1$, the quartic contribution reduces to a constant,
\begin{equation}
\Sigma^{ij}
C^{-1}_{ijkl}
\Sigma^{kl}
=
\frac{
3\mu(B_1+B_2)^2
+
4\lambda B_2^2
}{
\mu(3\lambda+2\mu)
}.
\end{equation}
Therefore \(\Delta\mathcal{S}\) only produces an overall shift of the ground-state energy and does not modify the magnetic dynamics. Consequently, the effective theory is fully determined by the gauge-field action \(\mathcal{S}_{A}\), the Kalb--Ramond coupling \(\mathcal{S}_{\mathrm{KR}}\), and the magnetoelastic gauge coupling \(\mathcal{S}_{A-m}\).

Throughout the Supplemental Material we employ the Fourier transform, and its inverse,
\begin{equation}
f({\bs r},t)=\int \frac{d\omega\, d{\bs q}}{(2\pi)^4}
\,e^{-i(\omega t-\bs q\cdot\bs r)}\,
f(\omega,\bs q), \qquad f(\omega,\bs q)=\int dt\, d{\bs r}\,
e^{i(\omega t-\bs q\cdot\bs r)}f({\bs r},t),
\end{equation}
respectively. For compactness we introduce the notation $q=(\omega,\mathbf q)$ and $\kappa=(\omega,q_z)$, where \(\kappa\) denotes the reduced momentum-frequency variable appropriate for quasi-one-dimensional propagation along the dislocation line. Thus the integration measures are defined as $d^4q\equiv{d\omega\, d^3q}/{(2\pi)^4}$
and $d\kappa\equiv{d\omega\, dq_z}/{(2\pi)^2}$.

\subsection{Derivation of the propagator 
\(G_{\alpha\beta}^{IJ,ij}\)}\label{sec:propagator}
The elastic gauge propagator is defined as the inverse of the quadratic kernel, $G_{\alpha\beta}^{IJ,ij}(q)
=
\left[
H_{\alpha\beta}^{IJ,ij}(q)
\right]^{-1},$ where $\alpha,\beta=A,B$. The quadratic kernel components are:
\begin{align}
H_{AA}^{IJ,ij}(q)
&=
C^{-1}_{lIrJ}
\epsilon_{lai}\epsilon_{rbj}q_aq_b
-\frac{1}{\xi}q_iq_j\delta^{IJ},
\\
H_{AB}^{IJ,ij}(q)
&=
-\,\omega\,
C^{-1}_{iIjJ}
\epsilon_{laj}q_a,
\\
H_{BA}^{IJ,ij}(q)
&=
\omega\,
C^{-1}_{jJiI}
\epsilon_{lai}q_a,
\\
H_{BB}^{IJ,ij}(q)
&=
\omega^2 C^{-1}_{iIjJ}
-
\left[
\frac{q^2}{\xi}\delta_{ij}
+
\left(1-\frac1{\xi}\right)q_iq_j
\right]\delta^{IJ}.
\end{align}
For an isotropic elastic medium,
\begin{equation}
C^{-1}_{iIjJ}
=
-\frac{\lambda}{2\mu(3\lambda+2\mu)}\,\delta_{iI}\delta_{jJ}
+
\frac{1}{4\mu}\left(
\delta_{ij}\delta_{IJ}
+
\delta_{iJ}\delta_{Ij}
\right).
\end{equation}
The inverse of a block matrix gives the four propagator sectors.
Define the Schur complement $S_A=H_{AA}-H_{AB}H_{BB}^{-1}H_{BA}$. Then in explicit tensor notation this means
\begin{align}
G_{AA}^{IJ,ij}
&=
\left[
S_A^{-1}
\right]^{IJ,ij},
\\
G_{AB}^{IJ,ij}
&=
-
\left[
S_A^{-1}
\right]^{IK,ik}
H_{AB}^{KL,kl}
\left[
H_{BB}^{-1}
\right]^{LJ,lj},
\\
G_{BA}^{IJ,ij}
&=
-
\left[
H_{BB}^{-1}
\right]^{IK,ik}
H_{BA}^{KL,kl}
\left[
S_A^{-1}
\right]^{LJ,lj},
\\
G_{BB}^{IJ,ij}
&=
\left[
H_{BB}^{-1}
\right]^{IJ,ij}
+
\left[
H_{BB}^{-1}
\right]^{IK,ik}
H_{BA}^{KL,kl}
\left[
S_A^{-1}
\right]^{MN,mn}
H_{AB}^{NP,np}
\left[
H_{BB}^{-1}
\right]^{PJ,pj},
\end{align}
where repeated indices are summed over, $K,L,M,N,P=x,y,z$ and $k,l,m,n,p=x,y,z$. Thus the full tensor propagator is
\begin{equation}
G_{\alpha\beta}^{IJ,ij}
=
\begin{pmatrix}
G_{AA}^{IJ,ij} & G_{AB}^{IJ,ij}\\
G_{BA}^{IJ,ij} & G_{BB}^{IJ,ij}
\end{pmatrix}
\end{equation}
with each component determined by the expressions above. Equipped with this result, we are in position to find the mixed magnon--dislon vertex obtained after integrating out the elastic gauge field:
\begin{equation}
V^{ab}(\kappa;b)
=
b_I
\Big[
q_z^2G_{AA}^{Iz,ac}\epsilon_{cb}
-\omega^2\epsilon_{ac}
\left(
G_{BB}^{Iz,cb}
+
G_{BB}^{Ib,cz}
\right)
-\omega q_z
\left(
G_{AB}^{Ib,az}
+
G_{AB}^{Iz,ab}
-
\epsilon_{ac}G_{BA}^{Iz,cd}\epsilon_{db}
\right)
\Big],
\label{eq:generalV}
\end{equation}
where \(a,b,c,d=x,y\), \(\kappa=(\omega,q_z)\), and repeated transverse
indices are summed.
\\
\\
{\bf Screw-dislocations:} according with the geometry under consideration $\bs b=b\hat z$ and $b_I=b\delta_{Iz}$. Therefore only the component $I=z$ contributes to Eq.~\eqref{eq:generalV}.
Thus the vertex becomes
\begin{equation}
V_{\rm screw}^{ab}(\kappa)=b
\Big[
q_z^2G_{AA}^{zz,ac}\epsilon_{cb}
-\omega^2\epsilon_{ac}
\left(
G_{BB}^{zz,cb}
+
G_{BB}^{zb,cz}
\right)
-\omega q_z
\left(
G_{AB}^{zb,az}
+
G_{AB}^{zz,ab}
-
\epsilon_{ac}G_{BA}^{zz,cd}\epsilon_{db}
\right)
\Big].
\label{eq:VscrewStart}
\end{equation}
Later, in the strict quasi-one-dimensional limit, ${\bs q}=q_z\hat z$ and $q_\perp=0$,
the transverse isotropic propagators have the form
\begin{align}
G_{AA}^{zz,ab}
&=
G_{AA}^{T}\delta^{ab}, \qquad G_{AB}^{zz,ab}=G_{AB}^{T}\epsilon^{ab},
\\
G_{BB}^{zz,ab}
&=
G_{BB}^{T}\delta^{ab}, \qquad G_{BA}^{zz,ab}=G_{BA}^{T}\epsilon^{ab}.
\end{align}
All propagator components with one transverse Burgers index and one longitudinal Burgers index vanish in the isotropic \(q_\perp=0\) limit, $G_{BB}^{zb,cz}=G_{AB}^{zb,az}=0$. Thus Eq.~\eqref{eq:VscrewStart} reduces to
\begin{equation}
V_{\rm screw}^{ab}(\kappa)=b\Big[
q_z^2G_{AA}^{T}\delta^{ac}\epsilon_{cb}-\omega^2\epsilon_{ac}G_{BB}^{T}\delta^{cb}
-\omega q_z
\left(
G_{AB}^{T}\epsilon^{ab}-\epsilon_{ac}G_{BA}^{T}\epsilon^{cd}\epsilon_{db}
\right)\Big].
\end{equation}
Using the identities $\delta^{ac}\epsilon_{cb}=\epsilon^a_{\ b}$, $\epsilon_{ac}\delta^{cb}=\epsilon_a^{\ b}$, and $\epsilon_{ac}\epsilon^{cd}\epsilon_{db}=-\epsilon_{ab},$
and writing all transverse antisymmetric tensors with the same index
convention, we obtain
\begin{equation}
V_{\rm screw}^{ab}(\kappa)
=
b
\left[
q_z^2G_{AA}^{T}
-\omega^2G_{BB}^{T}
-\omega q_z
\left(
G_{AB}^{T}
+
G_{BA}^{T}
\right)
\right]\epsilon^{ab}.
\label{eq:VscrewRT}
\end{equation}
Equivalently, defining the transverse gauge response, $R_T(\kappa)\equiv q_z^2G_{AA}^{T}
-\omega q_z
\left(
G_{AB}^{T}
+
G_{BA}^{T}
\right)
-\omega^2G_{BB}^{T}$, the screw vertex becomes, $V_{\rm screw}^{ab}(\kappa)=b\,R_T(\kappa)\,\epsilon^{ab}.$ For an isotropic elastic medium the transverse response evaluates to
\begin{equation}
R_T(\kappa)
=
\frac{2c_T^2\omega^2}
{c_T^2q_z^2-\omega^2}.
\label{eq:RTresult}
\end{equation}
This quantity determines both the magnon self-energy and the
magnon--dislon coupling. Therefore, one obtains
\begin{equation}
V_{\rm screw}^{ab}(\kappa)=\frac{2bc_T^2\omega^2}
{c_T^2q_z^2-\omega^2}
\,\epsilon^{ab}.
\end{equation}
\\
\\
{\bf Edge-dislocations:} for the considered geometry, the Burger vector is $\bs b=b\hat x$ and $b_I=b\,\delta_{Ix}$. Therefore, the vertex Eq. (\ref{eq:generalV}) becomes
\begin{equation}
V_{\rm edge}^{ab}(\kappa)=b\Big[
q_z^2G_{AA}^{xz,ac}\epsilon_{cb}
-\omega^2\epsilon_{ac}
\left(
G_{BB}^{xz,cb}
+
G_{BB}^{xb,cz}
\right)
-\omega q_z
\left(
G_{AB}^{xb,az}
+
G_{AB}^{xz,ab}
-
\epsilon_{ac}G_{BA}^{xz,cd}\epsilon_{db}
\right)
\Big].
\label{eq:Vedge_start}
\end{equation}
Let's consider the quasi-one-dimensional
approximation $\bs q=q_z\hat z$ and $q_\perp=0$.
It is useful to define $E_{ia}\equiv \epsilon_{iza}$ with $a=x,y$. Then, $E_{za}=0$ and $E_{xa},E_{ya}\neq 0$. First, consider the component $H_{AA}$,
\begin{equation}
H_{AA}^{xz,ac}=q_z^2
C^{-1}_{lxrz}E_{la}E_{rc}
-\frac1{\xi}q_aq_c\delta_{xz}=0,
\end{equation}
since $C^{-1}_{lxrz}=C_1\,\delta_{lx}\delta_{rz}+C_2\,\delta_{lz}\delta_{xr}$. The same analysis apply for the other components:
\begin{align}
H_{BB}^{xz,cb}
&=
\omega^2 C^{-1}_{cx bz}
-
\left[
\frac{q_z^2}{\xi}\delta_{cb}
+
\left(1-\frac1{\xi}\right)q_cq_b
\right]\delta_{xz}=0\\
H_{BB}^{xb,cz}
&=
\omega^2 C^{-1}_{cxzb}
-
\left[
\frac{q_z^2}{\xi}\delta_{cz}
+
\left(1-\frac1{\xi}\right)q_cq_z
\right]\delta_{xb}=0.
\end{align}

Now consider the mixed $AB$ and $BA$ blocks,  $H_{AB}^{IJ,ij}$ and $H_{BA}^{IJ,ij}$, respectively. In the strict \(q_\perp=0\) limit, $H_{AB}^{xb,az}=H_{AB}^{xz,ab}=0$ and $H_{BA}^{xz,cd}=0$. Therefore the
corresponding propagator components vanish, $G_{AA}^{xz,ac}=
G_{BB}^{xz,cb}=
G_{BB}^{xb,cz}=
G_{AB}^{xb,az}=
G_{AB}^{xz,ab}=
G_{BA}^{xz,cd}=0.$ Therefore, $V_{\rm edge}^{ab}(\kappa)=0,$ in the strict quasi-one-dimensional isotropic approximation. The vanishing of Eq.~(\ref{eq:Vedge_start}) is therefore a selection rule
of the strict $q_\perp=0$ approximation. To obtain the first nonzero
edge-dislocation coupling, we must retain a small but finite transverse
momentum,
\begin{equation}
\bs q=(q_x,q_y,q_z),
\qquad
q_\perp=\sqrt{q_x^2+q_y^2}\ll |q_z|.
\end{equation}
We expand the gauge propagator as
\begin{equation}
G_{\alpha\beta}^{IJ,ij}(\omega,q_x,q_y,q_z)
=
G_{\alpha\beta}^{(0)\,IJ,ij}(\omega,q_z)
+
q_xG_{\alpha\beta}^{(x)\,IJ,ij}(\omega,q_z)
+
q_yG_{\alpha\beta}^{(y)\,IJ,ij}(\omega,q_z)
+
O(q_\perp^2),
\label{eq:G_expansion_qperp}
\end{equation}
where $G^{(0)}$ denotes the propagator evaluated at $q_x=q_y=0$.
Since $G=H^{-1}$, the linear correction follows from $G^{(\eta)}=-G^{(0)}H^{(\eta)}G^{(0)}$, with $\eta=x,y$,
where $H^{(\eta)}
=
\left.
{\partial H}/{\partial q_\eta}
\right|_{q_x=q_y=0}.$ In components,
\begin{equation}
G_{\alpha\beta}^{(\eta)\,IJ,ij}
=
-
G_{\alpha\gamma}^{(0)\,IK,ik}
H_{\gamma\delta}^{(\eta)\,KL,kl}
G_{\delta\beta}^{(0)\,LJ,lj},
\label{eq:G_linear_components}
\end{equation}
where repeated indices are summed over
$\gamma,\delta=A,B$. The derivatives of the quadratic kernel are
\begin{align}
H_{AA}^{(\eta)\,IJ,ij}
&=
q_z C^{-1}_{lIrJ}
\left(
\epsilon_{l\eta i}\epsilon_{rzj}
+
\epsilon_{lzi}\epsilon_{r\eta j}
\right)
-\frac{q_z}{\xi}
\left(
\delta_{i\eta}\delta_{jz}
+
\delta_{iz}\delta_{j\eta}
\right)\delta_{IJ},
\\
H_{AB}^{(\eta)\,IJ,ij}
&=
-\omega C^{-1}_{iIjJ}\epsilon_{l\eta j},
\\
H_{BA}^{(\eta)\,IJ,ij}
&=
\omega C^{-1}_{jJiI}\epsilon_{l\eta i},
\\
H_{BB}^{(\eta)\,IJ,ij}
&=
-\left(1-\frac1\xi\right)q_z
\left(
\delta_{i\eta}\delta_{jz}
+
\delta_{iz}\delta_{j\eta}
\right)\delta_{IJ}.
\end{align}
Substituting Eq.~(\ref{eq:G_expansion_qperp}) into the edge vertex gives
the leading finite-$q_\perp$ structure
\begin{equation}
V_{\rm edge}^{ab}(\omega,q_z,q_x,q_y)
=
q_xF_x^{ab}(\omega,q_z)
+
q_yF_y^{ab}(\omega,q_z)
+
O(q_\perp^2).
\label{eq:Vedge_F_expansion}
\end{equation}
The coefficient functions are
\begin{align}
F_\eta^{ab}(\omega,q_z)
=
b
\Big[
&\nonumber q_z^2G_{AA}^{(\eta)\,xz,ac}\epsilon_{cb}
-\omega^2\epsilon_{ac}
\left(
G_{BB}^{(\eta)\,xz,cb}
+
G_{BB}^{(\eta)\,xb,cz}
\right)
-\omega q_z
\left(
G_{AB}^{(\eta)\,xb,az}
+
G_{AB}^{(\eta)\,xz,ab}
-
\epsilon_{ac}G_{BA}^{(\eta)\,xz,cd}\epsilon_{db}
\right)
\Big].
\label{eq:F_eta_compact}
\end{align}
Using Eq.~(\ref{eq:G_linear_components}), this may be written entirely in
terms of the $q_\perp=0$ propagator and the first derivative of the
inverse propagator:
\begin{align}
F_\eta^{ab}(\omega,q_z)
=
-b
\Big[
&q_z^2
G_{A\gamma}^{(0)\,xK,ak}
H_{\gamma\delta}^{(\eta)\,KL,kl}
G_{\delta A}^{(0)\,Lz,lc}
\epsilon_{cb}
\\
&-\omega^2\epsilon_{ac}
\left(
G_{B\gamma}^{(0)\,xK,ck}
H_{\gamma\delta}^{(\eta)\,KL,kl}
G_{\delta B}^{(0)\,Lz,lb}
+
G_{B\gamma}^{(0)\,xK,ck}
H_{\gamma\delta}^{(\eta)\,KL,kl}
G_{\delta B}^{(0)\,Lb,lz}
\right)
\nonumber\\
&-\omega q_z
\left(
G_{A\gamma}^{(0)\,xK,ak}
H_{\gamma\delta}^{(\eta)\,KL,kl}
G_{\delta B}^{(0)\,Lb,lz}
+
G_{A\gamma}^{(0)\,xK,ak}
H_{\gamma\delta}^{(\eta)\,KL,kl}
G_{\delta B}^{(0)\,Lz,lb}
\right.
\nonumber\\
&\left.
\hspace{3.3cm}
-
\epsilon_{ac}
G_{B\gamma}^{(0)\,xK,ck}
H_{\gamma\delta}^{(\eta)\,KL,kl}
G_{\delta A}^{(0)\,Lz,ld}
\epsilon_{db}
\right)
\Big].
\label{eq:F_eta_explicit}
\end{align}
Eqs.~(\ref{eq:Vedge_F_expansion})--(\ref{eq:F_eta_explicit})
show that the first nonvanishing edge-dislocation vertex is linear in the
transverse momentum. The functions $F_x^{ab}$ and $F_y^{ab}$ encode the
anisotropic response associated with transverse magnon modes carrying
momenta along the Burgers-vector direction and perpendicular to it,
respectively. In particular, the homogeneous transverse magnon mode
$(n,m)=(0,0)$ remains dark, while the first transverse modes
$(n,m)=(1,0)$ and $(0,1)$ acquire finite couplings through the terms
proportional to $q_xF_x^{ab}$ and $q_yF_y^{ab}$.

\subsection{Magnetization--dislocation interaction}\label{sec:action-coupling}
After integrating out the elastic gauge field, the effective interaction between the magnetization and dislocation degrees of freedom takes the current--current form
\begin{equation}
S_c=\int d^4q\,({\cal J}_\alpha)^i_I(-q)\,
G^{IJ}_{\alpha\beta,ij}(q)\,
({\cal J}_\beta)^j_J(q),
\label{eq:Sc_general}
\end{equation}
where \(G^{IJ}_{\alpha\beta,ij}(q)\) is the elastic gauge propagator. The coupling decomposes as $S_c=S_c^{mm}+S_c^{RR}+S_c^{mR}$, being the total current decomposed as ${\cal J}_\alpha={\cal J}_{m,\alpha}+{\cal J}_{d,\alpha}$. Here, the temporal and spatial sectors are defined by $({\cal J}_A)^i_I={\cal J}^{i0}_I$ and $({\cal J}_B)^i_I=\epsilon^{ijk}{\cal J}^{jk}_I/2$. Thus, we find
\begin{align}
S_c^{mm}&=\int d^4q\,({\cal J}_{m,\alpha})^i_I(-q)\,G^{IJ}_{\alpha\beta,ij}(q)\,({\cal J}_{m,\beta})^j_J(q),
\\
S_c^{RR}&=\int d^4q\,({\cal J}_{d,\alpha})^i_I(-q)\,
G^{IJ}_{\alpha\beta,ij}(q)\,({\cal J}_{d,\beta})^j_J(q),
\\
S_c^{mR}&=\int d^4q\,\left[({\cal J}_{m,\alpha})^i_I(-q)\,G^{IJ}_{\alpha\beta,ij}(q)\,({\cal J}_{d,\beta})^j_J(q)+({\cal J}_{d,\alpha})^i_I(-q)\,G^{IJ}_{\alpha\beta,ij}(q)\,({\cal J}_{m,\beta})^j_J(q)\right].
\label{eq:SmR_general}
\end{align}

In the static limit, $\partial_t m_i=0$ and $\partial_t R_i=0$, the \(B\)-sector currents vanish, $({\cal J}_{m,B})^i_I=({\cal J}_{d,B})^i_I=0$. Thus only the \(AA\) component of the propagator contributes. The mixed interaction becomes
\begin{align}
S_{c,\mathrm{stat}}^{mR}=\int d^3r\,d^3r'\,
\Big[&({\cal J}_{m,A})^i_I(\bs r)\,
G^{IJ}_{AA,ij}(\bs r-\bs r')\,
({\cal J}_{d,A})^j_J(\bs r')+({\cal J}_{d,A})^i_I(\bs r)\,
G^{IJ}_{AA,ij}(\bs r-\bs r')\,
({\cal J}_{m,A})^j_J(\bs r')
\Big].
\label{eq:SmR_static_general}
\end{align}
Using the reciprocity relation $G^{IJ}_{AA,ij}(\bs r-\bs r')=G^{JI}_{AA,ji}(\bs r'-\bs r)$ the two terms in Eq.~\eqref{eq:SmR_static_general} are equivalent. Therefore,
\begin{equation}
S_{c,\mathrm{stat}}^{mR}=2\int d{\bs r}\,d{\bs r'}\,({\cal J}_{m,A})^i_I(\bs r)\,
G^{IJ}_{AA,ij}(\bs r-\bs r')\,
({\cal J}_{d,A})^j_J(\bs r').
\label{eq:SmR_static_symmetric}
\end{equation}
For an isotropic elastic solid, the magnetic current is $({\cal J}_{m,A})^i_I=2B_2\mu^{-1}\epsilon^{ilk}\left(m_l\partial_k m_I+m_I\partial_k m_l
\right)$, while for a straight static dislocation oriented along the \(z\)-axis, $({\cal J}_{d,A})^j_J(\bs r')=b_J\delta^j_z\delta^{(2)}(\bs r'_\perp)$. Substituting those expressions into Eq.~\eqref{eq:SmR_static_symmetric} gives
\begin{align}
S_{c,\mathrm{stat}}^{mR}&\nonumber=\frac{4B_2}{\mu}
\int d{\bs r}\,d{\bs r'}\,b_J\,
G^{IJ}_{AA,iz}(\bs r-\bs r')\,
\delta^{(2)}(\bs r'_\perp)
\epsilon^{ilk}\left(m_l\partial_k m_I+m_I\partial_k m_l\right)\\
&=\int d{\bs r}\,D^{Ilk}(\bs r_\perp)
\left(m_l\partial_k m_I+m_I\partial_k m_l
\right),
\label{eq:SmR_static_explicit}
\end{align}
where
\begin{equation}
D^{Ilk}(\bs r_\perp)=\frac{4B_2b_J}{\mu}\epsilon^{ilk}\int dz'\,G^{IJ}_{AA,iz}(\bs r_\perp,-z').
\label{eq:D_static}
\end{equation}
This term is the static limit of the general magnon--dislocation coupling and is determined by the dislocation geometry through the static elastic gauge propagator.}

\subsection{Coupled theory for magnons and dislons}\label{sec:magnon-dislons}
{We derive the low-energy effective theory for the coupled dynamics of magnons and dislons after integrating out the elastic gauge field. We expand around a ferromagnetic ground state polarized along the \(z\)-axis, ${\bs m}=\hat{\bs z}+\delta\bs m$, $\delta\bs m=(\delta m_x,\delta m_y,0)$ and $|\delta\bs m|\ll1$. We focus on quasi-one-dimensional propagation along a straight dislocation line, $q=(\omega,0,0,q_z)$. A straight dislocation oriented along \(z\) is parametrized as $R^\mu(t,z)=(t,X_1(t,z),X_2(t,z),z)^T$, where \(X_1\) and \(X_2\) are the transverse dislon fields. Within the basis of circular variables $\delta m_\pm=\delta m_x\pm i\delta m_y$ and $X_\pm=X_1\pm iX_2$. The quadratic action can then be written as $\mathcal S_{\rm eff}^{(2)}=\mathcal S_m^{(2)}+\mathcal S_R^{(2)}+\mathcal S_c^{(2)}$.

\subsubsection{Magnon sector}
The quadratic magnon action is,
\begin{equation}
\mathcal S_m^{(2)}=\int d\kappa\,
\begin{pmatrix}
\delta m_+^* &
\delta m_-^*
\end{pmatrix}
\Lambda_m(\kappa)
\begin{pmatrix}
\delta m_+\\
\delta m_-
\end{pmatrix},
\end{equation}
where $\Lambda_m(\kappa)=\omega\sigma_z+\zeta_\kappa\sigma_x-\omega_m(q_z)\sigma_0$, with the bare magnon dispersion $\omega_m(q_z)=\omega_0+Jq_z^2$, $J={2A\gamma}/{M_s}$, and $\omega_0=\gamma
\left(\mu_0H+{2K}/{M_s}\right)$. The magnetic current--current interaction generates a correction to the
quadratic magnon action proportional to the transverse response
\( R_T(\kappa)\). Here $\zeta_\kappa$ is the gauge-mediated magnon self-energy and explicitly given by $\zeta_\kappa=(2B_2\mu^{-1})^2
 R_T(\kappa)$, which, using Eq.~(\ref{eq:RTresult}), yields
\begin{equation}
\zeta_\kappa=\left(\frac{2B_2}{\mu}\right)^2\frac{2c_T^2\omega^2}{c_T^2q_z^2-\omega^2}.
\end{equation}
Thus the magnon self-energy is resonantly enhanced near the transverse
phonon pole \(\omega=c_T|q_z|\).

\subsubsection{Dislon sector}
The bare quadratic action for transverse dislocation fluctuations is
\begin{equation}
\mathcal S_R^{(2)}
=
\frac12
\int d\kappa\,
X_a(-\kappa)
\left[
\rho_0\omega^2
-
T_0q_z^2
\right]
\delta_{ab}
X_b(\kappa).
\end{equation}
The elastic gauge field generates an additional dislon self-energy,
\begin{equation}
\mathcal S_c^{RR,(2)}
=
\int d\kappa\,
X_a(-\kappa)
\Pi_{RR}^{ab}(\kappa)
X_b(\kappa),
\end{equation}
with
\begin{align}
\Pi_{RR}^{ab}(\kappa)
=
b_Ib_J
\Big[
q_z^2G_{AA}^{IJ,ab}
+
\omega q_z
\left(
G_{AB}^{IJ,ac}\epsilon_{cb}
+
\epsilon_{ac}G_{BA}^{IJ,cb}
\right)
+
\omega^2
\epsilon_{ac}G_{BB}^{IJ,cd}\epsilon_{db}
\Big].
\end{align}
The full dislon kernel is therefore $K_{RR}^{ab}(\kappa)
=
\left(
\rho_0\omega^2-T_0q_z^2
\right)\delta_{ab}
+
2\Pi_{RR}^{ab}(\kappa)$. 

For a screw dislocation, \(\mathbf b=b\hat{\mathbf z}\), cylindrical symmetry around the line implies $K_{RR,{s}}^{ab}=\left[\rho_{ s}(q_z)\omega^2-T_{ s}(q_z)q_z^2\right]\delta^{ab}$, where
\begin{align}
\rho_{s}(q_z)&=\rho_0-\frac{\rho_{ el}b^2}{4\pi}\ln(q_z^2L^2),\\
T_{ s}(q_z)&=T_0+\frac{\mu b^2}{\pi}
\left(1-\frac{c_T^2}{c_L^2}
\right)\ln(q_z^2L^2).
\end{align}
Thus, the screw-dislon dispersion is
\begin{equation}
\omega_{d,{s}}(q_z)=|q_z|
\sqrt{\frac{T_{ s}(q_z)}{\rho_{s}(q_z)}
},
\end{equation}
with the two circular polarizations \(X_\pm\) remaining degenerate.

For an edge dislocation, \(\mathbf b=b\hat{\mathbf x}\), cylindrical symmetry is explicitly broken. The transverse polarizations generally mix, and the dislon kernel is no longer proportional to \(\delta^{ab}\). The logarithmically renormalized coefficients are
\begin{align}
\rho_{ e}(q_z)
&=
\rho_0
-
\frac{\rho_{ el}b^2}{4\pi}
\ln(q_z^2L^2),\\
T_{ e}(q_z)
&=
T_0+
\frac{3\mu b^2}{16\pi}
\ln(q_z^2L^2).
\end{align}

Before glide projection, the edge sector is described by the full
two-polarization kernel. The physical low-energy theory is
obtained after projecting onto the glide-allowed mode,
\begin{equation}
\omega_{d,{ e}}(q_z)=|q_z|\sqrt{
\frac{T_{ e}(q_z)
}{\rho_{ e}(q_z)}
}.
\end{equation}
Integrating out the bulk elastic
degrees of freedom generates the self-energy is \(\Pi_{RR}^{ab}\), which renormalizes both the effective mass density and line tension. In the bare elastic-string limit,
\begin{equation}
T_0
\simeq
\frac{\mu b^2}{4\pi}
\ln\!\left(\frac{L}{a}\right),
\qquad
\rho_0
\simeq
\frac{\rho_{ el}b^2}{4\pi}
\ln\!\left(\frac{L}{a}\right),
\end{equation}
where \(a\) is the dislocation-core radius and \(L\) is an infrared
cutoff associated with the dislocation length. Consequently, $v_d=\sqrt{{T_0}/{\rho_0}}=c_T$,
showing that the bare dislon propagates at the transverse sound velocity.}

\subsubsection{Total magnon--dislon theory}
The quadratic interaction between magnons and dislons is given by
\begin{equation}
\mathcal S_c^{mR,(2)}
=\frac{2B_2}{\mu}\int d\kappa
\left[\delta m_a(-\kappa)V^{ad}(\kappa)X_d(\kappa)+{\rm h.c.}
\right],
\end{equation}
where the vertex is
\begin{align}
V^{ad}(\kappa)=b_I
\Big[q_z^2G_{AA}^{Iz,ac}\epsilon_{cd}-\omega^2\epsilon_{ac}\left(G_{BB}^{Iz,cd}+G_{BB}^{Id,cz}\right)-\omega q_z\left(G_{AB}^{Id,az}
+G_{AB}^{Iz,ad}-\epsilon_{ac}G_{BA}^{Iz,ce}\epsilon_{ed}
\right)\Big].
\label{eq:mixed_vertex_general}
\end{align}

Thus the coupled magnon--dislon theory has the compact form
\begin{equation}
\mathcal S_{\rm eff}
=
\int d\kappa\,
\Psi_\kappa^\dagger
\Lambda_\kappa
\Psi_\kappa,
\end{equation}
where $\Psi_\kappa=(\delta m_+,\delta m_-,X_+,X_-)^{T}$ and
\begin{equation}
\Lambda_\kappa=
\begin{pmatrix}
\Lambda_{m,\kappa} & \Xi_\kappa
\\
\Xi^\dagger_\kappa & \Lambda_{d,\kappa}
\end{pmatrix}.
\end{equation}

\subsection{Screw-dislocation hybridization and helicity selection}\label{sec:screw-dislons}
For a screw dislocation, the Burgers vector is parallel to the dislocation line, $\bs b=b\hat{\bs z}$. In this geometry the system preserves cylindrical symmetry around the $z$ axis. Consequently, the transverse response is isotropic in the $xy$ plane and only the $I=z$ sector of the gauge propagator contributes to the mixed magnon--dislon vertex. The vertex reduces to the antisymmetric form
\begin{equation}
V_{ s}^{ad}(\kappa)=b\,
\frac{2c_T^2\omega^2}
{c_T^2q_z^2-\omega^2}
\epsilon_{ad},
\label{eq:screw_vertex}
\end{equation}
where $a,d=x,y$. In the basis of circular variables, the quadratic kernel for screw magnon--dislon hybridization can be written as
\begin{equation}
\Lambda^{s}_{\kappa}
=
\begin{pmatrix}
\omega-\omega_m & \zeta_\kappa & 0 & -i\Gamma^{s}_{\kappa}
\\
\zeta_\kappa &
-\omega-\omega_m &
i\Gamma^{s}_{\kappa} &
0
\\
0 &
-i\Gamma^{s}_{\kappa} &
D_{ s} &
0
\\
i\Gamma^{s}_{\kappa} &
0 &
0 &
D_{ s}
\end{pmatrix},
\label{eq:screw_kernel}
\end{equation}
where $D_{ s}(\omega,q_z)
=\rho_{ s}(q_z)\omega^2
-T_{ s}(q_z)q_z^2
\equiv
\Lambda_{d,\kappa}^{s}.$  The structure of Eq.~\eqref{eq:screw_kernel} makes the helicity selection rule explicit. The positive-helicity magnon $\delta m_+$ couples only to the opposite circular dislon polarization $X_-$, while $\delta m_-$ couples only to $X_+$. This selection rule follows from conservation of angular momentum around the screw-dislocation axis. Since the positive-frequency magnon sector is represented by $\delta m_+$, the resonant hybridization occurs in the subspace $\Phi_{\rm B}
= (\delta m_+,X_-)^T$. The orthogonal dislon polarization $X_+$ carries no magnon spectral weight in this sector and remains a dark screw-dislon mode. Neglecting the off-resonant negative-frequency magnon $\delta m_-$, the bright resonant kernel becomes
\begin{equation}
\Lambda_{\rm B}(\omega,q_z)
=
\begin{pmatrix}
\omega-\omega_m(q_z) & -i\Gamma^{s}_{\kappa}
\\
i\Gamma^{s}_{\kappa} & D_{s}(\omega,q_z)
\end{pmatrix}.
\label{eq:screw_bright_kernel}
\end{equation}
The hybrid poles satisfy
\begin{equation}
\det \Lambda_{\rm B}=\left[\omega-\omega_m(q_z)\right]
D_{s}(\omega,q_z)-\left|\Gamma^{s}_{\kappa}\right|^2=0.
\label{eq:screw_det_bright}
\end{equation}
Close to the dislon pole,
\begin{equation}
D_{ s}(\omega,q_z)
\simeq
Z_d(q_z)\left[\omega-\omega_{d,s}(q_z)\right],
\qquad
Z_d(q_z)
=
\left.
\frac{\partial D_{ s}}{\partial \omega}
\right|_{\omega=\omega_{d,s}(q_z)}
\simeq
2\rho_{ s}(q_z)\omega_{d,s}(q_z),
\end{equation}
neglecting the weak frequency dependence of $\rho_{ s}$ and $T_s$ 
Eq.~\eqref{eq:screw_det_bright} then reduces to an effective two-level problem,
\begin{equation}
\left[\omega-\omega_m(q_z)\right]
\left[\omega-\omega_{d,s}(q_z)\right]
-
g_{s}^2(q_z)
=
0,
\end{equation}
with the pole-normalized coupling
\begin{equation}
g_{s}(q_z)
=
\left.
\frac{
\left|\Gamma^{s}_{\kappa}\right|
}{
\sqrt{Z_d(q_z)}
}
\right|_{\omega=\omega_{d,s}(q_z)} .
\end{equation}
The resulting bright hybrid branches are
\begin{equation}
\omega_{\pm}(q_z)
=\frac{\omega_m(q_z)+\omega_{d,s}(q_z)}{2}\pm \frac12\sqrt{\left[\omega_m(q_z)-\omega_{d,s}(q_z)
\right]^2+4g_{s}^2(q_z)}.
\label{eq:screw_hybrid_spectrum}
\end{equation}
At resonance, $\omega_m(q_\ast)=\omega_{d,s}(q_\ast)$, the avoided-crossing gap is $\Delta_{ s}=2g_{ s}(q_\ast).$ The corresponding normalized bright eigenstates may be written as
\begin{equation}
\Phi_\pm
=
\frac{1}{\mathcal N_\pm}
\begin{pmatrix}
1
\\
\dfrac{i\left[\omega_\pm-\omega_m(q_z)\right]}
{\Gamma^s_{\kappa}}
\end{pmatrix},
\end{equation}
where
\begin{equation}
\mathcal N_\pm^2=1+\left|
\frac{\omega_\pm-\omega_m(q_z)}{
\Gamma^s_{\kappa}}\right|^2 .
\end{equation}
Equivalently, after pole normalization of the dislon field, one may write $\Phi_\pm
=u_m^{(\pm)}\delta m_+
+u_d^{(\pm)}X_-$, with $\left|u_m^{(\pm)}\right|^2+
\left|u_d^{(\pm)}\right|^2=1$. The magnon and dislon spectral weights are therefore
\begin{equation}
Z_m^{(\pm)}
=
\left|u_m^{(\pm)}\right|^2,
\qquad
Z_d^{(\pm)}
=
\left|u_d^{(\pm)}\right|^2 .
\end{equation}
These weights are exchanged across the avoided crossing, as expected for a two-level hybridization problem. The dark screw-dislon mode satisfies $D_{s}(\omega,q_z)=0,$
and hence has dispersion $\omega_{\rm dark}(q_z)=\omega_{d,s}(q_z).$ Because this mode is proportional to $X_+$ in the positive-frequency magnon sector, it does not couple to $\delta m_+$ and carries no magnon spectral weight.

Finally, the screw geometry preserves circular polarization. The hybrid bright modes inherit the circular character of the participating magnon and dislon helicities. Therefore, unlike the edge case, screw-dislocation hybridization does not generate ellipticity in the transverse spin precession:
\begin{equation}
\mathcal E_{ s}
=
\frac{
|\delta m_x|^2-|\delta m_y|^2
}{
|\delta m_x|^2+|\delta m_y|^2
}
=
0.
\end{equation}
This contrasts directly with the edge-dislocation case, where glide polarization breaks cylindrical symmetry and produces finite ellipticity.

\subsection{Hybridization between edge dislons and transverse magnon modes}\label{sec:edge-dislons}
For an edge dislocation, $\bs b=b\hat{\bs x}$, the mixed vertex takes the form
\begin{align}
V_{ e}^{ab}(\kappa)=b\Big[
&q_z^2G_{AA}^{xz,ac}\epsilon_{cb}
-\omega^2\epsilon_{ac}
\left(
G_{BB}^{xz,cb}
+
G_{BB}^{xb,cz}
\right)
-\omega q_z
\left(
G_{AB}^{xb,az}
+
G_{AB}^{xz,ab}
-
\epsilon_{ac}G_{BA}^{xz,cd}\epsilon_{db}
\right)
\Big].
\label{eq:edge_vertex}
\end{align}
Unlike the screw geometry, the edge vertex vanishes identically in the strict quasi-one-dimensional limit $q_\perp=0$. Consequently, the homogeneous magnon mode does not hybridize with the glide-projected edge dislon. The leading coupling therefore arises from finite transverse momentum and is generated by the first transverse magnon harmonics $\psi^{10}$ and $\psi^{01}$. Expanding the vertex around $q_\perp=0$ yields
\begin{equation}
V_{ e}^{ab}=q_xF_x^{ab}(\omega,q_z)+q_yF_y^{ab}(\omega,q_z)
+O(q_\perp^2),
\end{equation}
which induces the couplings
\begin{align}
\Gamma_{+g}^{10}
&=
\frac{2B_2}{\mu}
\frac{\pi}{L_x}
\left(
F_x^{xx}
-
iF_x^{yx}
\right),
\\
\Gamma_{+g}^{01}
&=
\frac{2B_2}{\mu}
\frac{\pi}{L_y}
\left(
F_y^{xx}
-
iF_y^{yx}
\right).
\end{align}
For the glide polarization $e_g=\hat{\bs x}$, the relevant components are $F_x^{xx}$, $F_x^{yx}$, $F_y^{xx}$, and $F_y^{yx}$. They are obtained from the finite-$q_\perp$ derivative of the full gauge propagator,
\begin{equation}
G_{\alpha\beta}^{(\eta)\,IJ,ij}(\omega,q_z)
\equiv
\left.
\frac{\partial G_{\alpha\beta}^{IJ,ij}(\omega,\bs q)}
{\partial q_\eta}
\right|_{q_x=q_y=0},
\end{equation}
with $\eta=x,y$. Explicitly,
\begin{align}
F_\eta^{xx}
&=
b\Big[
-q_z^2G_{AA}^{(\eta)\,xz,xy}
-\omega^2
\left(
G_{BB}^{(\eta)\,xz,yx}
+
G_{BB}^{(\eta)\,xx,yz}
\right)
-\omega q_z
\left(
G_{AB}^{(\eta)\,xx,xz}
+
G_{AB}^{(\eta)\,xz,xx}
+
G_{BA}^{(\eta)\,xz,yy}
\right)
\Big],
\\
F_\eta^{yx}
&=
b\Big[
-q_z^2G_{AA}^{(\eta)\,xz,yy}
+
\omega^2
\left(
G_{BB}^{(\eta)\,xz,xx}
+
G_{BB}^{(\eta)\,xx,xz}
\right)
-\omega q_z
\left(
G_{AB}^{(\eta)\,xx,yz}
+
G_{AB}^{(\eta)\,xz,yx}
-
G_{BA}^{(\eta)\,xz,xy}
\right)
\Big].
\end{align}
For the isotropic elastic kernel used in the main text, these components reduce to
\begin{align}
F_x^{xx}(\omega,q_z)&=0,
\\
F_x^{yx}(\omega,q_z)
&=
-\frac{
b\mu q_z\,\mathcal P_x(\omega,q_z)
}{
(\mu q_z^2-\omega^2)\,
\mathcal Q_x(\omega,q_z)
},
\\
F_y^{xx}(\omega,q_z)
&=
\frac{b\mu}{q_z},
\\
F_y^{yx}(\omega,q_z)&=0,
\end{align}
where
\begin{align}
\mathcal P_x(\omega,q_z)
=&
-3\lambda^2\mu q_z^2\omega^2
+3\lambda^2\omega^4
+12\lambda\mu^3q_z^4
-8\lambda\mu^2q_z^2\omega^2
+6\lambda\mu\omega^4
+8\mu^4q_z^4
-12\mu^3q_z^2\omega^2
+4\mu^2\omega^4,
\\
\mathcal Q_x(\omega,q_z)
=&\,
3\lambda^2\mu q_z^4
-3\lambda^2q_z^2\omega^2
+8\lambda\mu^2q_z^4
-7\lambda\mu q_z^2\omega^2
+\lambda\omega^4
+4\mu^3q_z^4
-6\mu^2q_z^2\omega^2
+2\mu\omega^4 .
\end{align}
The resulting finite-$q_\perp$ coupling is the leading symmetry-allowed interaction between transverse magnons and the glide edge dislon.

At low energies, the glide constraint removes one of the transverse dislon polarizations, leaving a single propagating mode $X_g$ aligned with the Burgers vector. Retaining only the first transverse magnon harmonics and the glide mode, the resonant sector is spanned by
\begin{equation}
\Phi=
\left(
\psi^{10},
\psi^{01},
X_g
\right)^T .
\end{equation}
In the generic case $L_x\neq L_y$, the transverse magnon modes are nondegenerate and the effective Hamiltonian becomes
\begin{equation}
H_{\rm eff}(q_z)
=
\begin{pmatrix}
\omega_{10}(q_z) & 0 & g_{10}
\\
0 & \omega_{01}(q_z) & g_{01}
\\
g_{10}^{*} & g_{01}^{*} & \omega_{d,e}(q_z)
\end{pmatrix},
\label{eq:Heff_edge}
\end{equation}
where $g_{10}=|\Gamma_{+g}^{10}|$ and $g_{01}=|\Gamma_{+g}^{01}|$. Although the microscopic vertices $\Gamma_{+g}^{10}$ and $\Gamma_{+g}^{01}$ inherit the frequency and momentum dependence of the functions $F_\eta^{ab}(\omega,q_z)$, we use an on-shell approximation and evaluate them at the corresponding bare resonance. This approximation is justified when the vertices vary slowly over the narrow hybridization window set by $|g_{10}|$ and $|g_{01}|$. We also neglect higher transverse harmonics and terms of order $O(q_\perp^2)$.

The hybrid eigenfrequencies are determined by
\begin{equation}
(\omega-\omega_{10})
(\omega-\omega_{01})
(\omega-\omega_d)
-
|g_{10}|^2(\omega-\omega_{01})
-
|g_{01}|^2(\omega-\omega_{10})
=
0 .
\label{eq:cubic_edge}
\end{equation}
The corresponding normalized eigenstates are written as
\begin{equation}
\Phi_n
=
\begin{pmatrix}
u_{10}^{(n)}
\\
u_{01}^{(n)}
\\
v_g^{(n)}
\end{pmatrix},
\end{equation}
where
\begin{equation}
u_{10}^{(n)}
=
\frac{g_{10}}{\omega_n-\omega_{10}}\,v_g^{(n)},
\qquad
u_{01}^{(n)}
=
\frac{g_{01}}{\omega_n-\omega_{01}}\,v_g^{(n)}.
\end{equation}
Equivalently,
\begin{equation}
\Phi_n
=
\frac{1}{\mathcal N_n}
\begin{pmatrix}
\dfrac{g_{10}}{\omega_n-\omega_{10}}
\\[0.8em]
\dfrac{g_{01}}{\omega_n-\omega_{01}}
\\[0.8em]
1
\end{pmatrix},
\end{equation}
with
\begin{equation}
\mathcal N_n^2
=
1+
\frac{|g_{10}|^2}{|\omega_n-\omega_{10}|^2}
+
\frac{|g_{01}|^2}{|\omega_n-\omega_{01}|^2}.
\end{equation}
The magnon and dislon spectral weights are therefore
\begin{equation}
Z_m^{(n)}
=
|u_{10}^{(n)}|^2
+
|u_{01}^{(n)}|^2,
\qquad
Z_d^{(n)}
=
|v_g^{(n)}|^2,
\end{equation}
and satisfy $Z_m^{(n)}+Z_d^{(n)}=1$.
Unlike screw dislons, which preserve circular polarization, edge dislons are glide polarized along the Burgers vector. The resulting hybridization transfers this anisotropy to the magnon sector and deforms the spin precession from circular to elliptical. The transverse spin amplitudes associated with the $n$-th hybrid mode are reconstructed from the projected edge vertices according to
\begin{align}
\delta m_x^{(n)}
&=
u_{10}^{(n)}F_x^{xx}
+
u_{01}^{(n)}F_y^{xx},
\\
\delta m_y^{(n)}
&=
u_{10}^{(n)}F_x^{yx}
+
u_{01}^{(n)}F_y^{yx}.
\end{align}
The resulting ellipticity is defined as
\begin{equation}
\mathcal E_n(q_z)
=
\frac{
|\delta m_x^{(n)}|^2
-
|\delta m_y^{(n)}|^2
}{
|\delta m_x^{(n)}|^2
+
|\delta m_y^{(n)}|^2
}.
\end{equation}
The limits $\mathcal E_n=0$ and $|\mathcal E_n|=1$ correspond to circular and linearly polarized oscillations, respectively. A finite ellipticity persists even away from resonance due to the intrinsic anisotropy of the edge vertex $F_\eta^{ab}$, while pronounced enhancements occur near the avoided crossings where magnon and glide-dislon components are maximally mixed.

Note that for \(L_x=L_y\), the degenerate transverse modes reorganize into bright and dark combinations; only the bright mode couples to the glide dislon, while the dark mode remains uncoupled.

\end{document}